\documentstyle[12pt]{article}

\begin{document}
\begin{titlepage}
\begin{center}

{\Large\bf The Importance of Higher Twist Corrections \\
\vskip 0.5cm
in Polarized DIS}
\end{center}
\vskip 2cm
\begin{center}
{\bf Elliot Leader}\\
{\it Imperial College London\\ Prince Consort Road, London SW7 2BW,
England }
\vskip 0.5cm
{\bf Aleksander V. Sidorov}\\
{\it Bogoliubov Theoretical Laboratory\\
Joint Institute for Nuclear Research, 141980 Dubna, Russia }
\vskip 0.5cm
{\bf Dimiter B. Stamenov \\
{\it Institute for Nuclear Research and Nuclear Energy\\
Bulgarian Academy of Sciences\\
Blvd. Tsarigradsko Chaussee 72, Sofia 1784, Bulgaria }}
\end{center}

\vskip 0.3cm
\begin{abstract}
\hskip -5mm
The higher twist corrections $h^N(x)/Q^2$ to the spin
dependent proton and neutron $g_1$ structure functions are
extracted from the world data on $g_1(x,Q^2)$ in a model
independent way and found to be non-negligible. Their role in
determining the polarized parton densities in the nucleon is
discussed. It is also considered how the results are influenced
by the recent JLab and HERMES/d inclusive DIS data.
\end{abstract}
\vskip 4.0cm
\begin{center}
{\it This talk was presented at DIS'2003 Workshop,
St. Petersburg, 23-27 April, 2003.}
\end{center}
\end{titlepage}

\newpage
\setcounter{page}{1}

\section{Introduction}

Spurred on by the famous EMC experiment \cite{EMC} at CERN in
1987, there has been a huge growth of interest in {\it polarized}
DIS experiments which yield more refined information about the
partonic structure of the nucleon, {\it i.e.}, how the nucleon
spin is divided up among its constituents, quarks and gluons.
Many experiments have been carried out at SLAC, CERN, DESY and
JLab to measure the longitudinal ($A_{\parallel}$) and transverse
($A_{\perp}$) asymmetries and to extract from them the
photon-nucleon asymmetries $A_1(x, Q^2)$ and $A_2(x, Q^2)$ as
well as the nucleon spin-dependent structure functions
$g_1(x,Q^2)$ and $g_2(x,Q^2)$.

There is, however, an important difference between the kinematic
regions of the unpolarized and polarized data sets. While in the
unpolarized case we can cut the low $Q^2$ and $W^2$ data in order
to eliminate the less known non-perturbative higher twist effects,
it is impossible to perform such a procedure for the present data
on the spin-dependent structure functions without loosing too
much information. So, to extract correctly the polarized parton
densities from the experimental data a special attention should
be paid to the higher twist (powers in $1/Q^2$) corrections to
the nucleon structure functions. Their role in determining the
polarized parton densities in the nucleon using different
approaches of QCD fits to the data is discussed in this talk. It
is also considered how the results are influenced by the recent
JLab \cite{JLab} and HERMES/d \cite{HERMESd} data.

\section{QCD treatment of $g_1(x, Q^2)$}

In QCD the spin structure function $g_1$ can be written in the
following form:
\begin{equation}
g_1(x, Q^2) = g_1(x, Q^2)_{\rm LT} + g_1(x, Q^2)_{\rm HT}~,
\label{g1QCD}
\end{equation}
where "LT" denotes the leading twist ($\tau=2$) contribution to
$g_1$, while "HT" denotes the contribution to $g_1$ arising from
QCD operators of higher twist, namely $\tau \geq 3$. In
(\ref{g1QCD}) we have dropped the nucleon target label N. The HT
power corrections (up to ${\cal O}(1/Q^2)$ terms) can be divided
in two parts:
\begin{equation}
g_1(x, Q^2)_{\rm HT}=h(x, Q^2)/Q^2 + h^{\rm TMC}(x, Q^2)/Q^2~,
\label{HTQCD}
\end{equation}
where $h^{\rm TMC}(x, Q^2)$ are the exactly calculable {\cite{TB}
kinematic target mass corrections and $h(x, Q^2)$ are the {\it
dynamical} higher twist ($\tau=3$ and $\tau=4$) corrections to
$g_1$, which are related to multi-parton correlations in the
nucleon. The latter are non-perturbative effects and cannot be
calculated without using models. $g_1(x, Q^2)_{\rm LT}$ in
(\ref{g1QCD}) is the well known pQCD expression and in NLO has
the form
\begin{equation}
g_1(x,Q^2)_{\rm pQCD}={1\over 2}\sum _{q} ^{N_f}e_{q}^2 [(\Delta
q +\Delta\bar{q})\otimes (1 + {\alpha_s(Q^2)\over 2\pi}\delta
C_q) +{\alpha_s(Q^2)\over 2\pi}\Delta G\otimes {\delta C_G\over
N_f}],
\label{g1partons}
\end{equation}
where $\Delta q(x,Q^2), \Delta\bar{q}(x,Q^2)$ and $\Delta
G(x,Q^2)$ are quark, anti-quark and gluon polarized densities in
the proton, which evolve in $Q^2$ according to the spin-dependent
NLO DGLAP equations. $\delta C(x)_{q,G}$ are the NLO
spin-dependent Wilson coefficient functions and the symbol
$\otimes$ denotes the usual convolution in Bjorken $x$ space.
$\rm N_f$ is the number of active flavors.

\section{QCD fits to the data and the role of higher twists}

Up to now, two approaches have been mainly used to extract the
polarized parton densities (PPD) from the world polarized DIS
data. According to the first \cite{GRSV,LSS2001} the leading twist
LO/NLO QCD expressions for the structure functions $g_1^N$ and
$F_1^N$ have been used in order to confront the data on
$A_1(\approx g_1/F_1)$ and $g_1/F_1$. It was shown
\cite{LSS2001,LomConf} that in this case the extracted from the
world data ``effective'' HT corrections $h^{A_1}(x)$ to $A_1$
\begin{equation}
A_1(x,Q^2)=(1+\gamma^2){g_1(x,Q^2)_{\rm LT}\over F_1(x,Q^2)_{\rm
LT}} + {h^{A_1}(x)\over Q^2} \label{A1HT}
\end{equation}
are negligible and consistent with zero within the errors,
$h^{A_1}(x) \approx 0$ (see Fig.1). This result has been confirmed
independently in \cite{GRSV}. In Fig. 1 are also shown (open
circles) our new results on the HT corrections to $A_1$ including in
the world data set \cite{EMC,world} the recent JLab \cite{JLab} and
HERMES \cite{HERMESd} data on $g_1/F_1$ for neutron and deuteron,
respectively. As seen from Fig. 1, due to the much more precise JLab
and HERMES new data, the HT corrections $h^{A_1}(x)$ to $A_1$ for
the neutron and deuteron targets are much better determined now
at large $x$ and better consistent with zero in this kinematic
region.

What follows from these results is that the higher twist
corrections to $g_1$ and $F_1$ compensate each other in the ratio
$g_1/F_1$ and the PPD extracted this way are less sensitive to
higher twist effects.\\

According to the second approach \cite{SMCBB}, $g_1/F_1$ and
$A_1$ data have been fitted using phenomenological
parametrizations of the experimental data for $F_2(x,Q^2)$ and
$R(x,Q^2)$ ($F_1$ has been replaced by the usually extracted from
unpolarized DIS experiments $F_2$ and $R$). Note that such a
procedure is equivalent to a fit to $(g_1)_{exp}$, but it is
more precise than the fit to the $g_1$ data themselves
actually presented by the experimental groups.
The point is that most of the experimental data on $g_1$
have been extracted from the $A_1$ and $g_1/F_1$ data using the
additional assumption that the ratio $g_1/F_1$ does not depend on
$Q^2$. Also, different experimental groups have used {\it
different} parametrizations for $F_2$ and $R$.

If the second approach is applied to the data, the
``effective higher twist'' contribution $h^{A_1}(x)/Q^2$ to
$A_1(g_1/F_1)$ is found \cite{GRSV} to be sizeable and important in
the fit [the HT corrections to $g_1$ cannot be compensate
because the HT corrections to $F_1(F_2$ and $R)$ are absorbed
by the phenomenological parametrizations of the data on $F_2$ and
$R$]. Therefore, to extract correctly the polarized parton
densities from the $g_1$ data, the HT corrections to $g_1$ have
to be taken into account. Note that a QCD fit to the data in this
case, keeping in $g_1(x,Q^2)_{QCD}$ only the leading-twist
expression (as it was done in \cite{SMCBB}), leads to some
"effective" parton densities which involve in themselves the HT
effects and therefore, are not quite correct.

Keeping in mind the discussion above we have analyzed the world
data \cite{EMC,world} on inclusive polarized DIS taking into
account the higher twist corrections to the nucleon structure
function $g_1^N(x, Q^2)$. In our fit to the data we have used the
following expressions for $g_1/F_1$ and $A_1$:
\begin{eqnarray}
\nonumber \left[{g_1^N(x,Q^2)\over F_1^N(x,
Q^2)}\right]_{exp}~&\Leftrightarrow&~ {{g_1^N(x,Q^2)_{\rm
LT}+h^N(x)/Q^2}\over F_2^N(x,Q^2)_{exp}}2x{[1+
R(x,Q^2)_{exp}]\over (1+\gamma^2)}~,\\
A_1^N(x,Q^2)_{exp}~&\Leftrightarrow&~{{g_1^N(x,Q^2)_{\rm LT}+
h^N(x)/Q^2}\over F_2^N(x,Q^2)_{exp}}2x[1+R(x,Q^2)_{exp}]~,
\label{g1F2Rht}
\end{eqnarray}
where $g_1^N(x,Q^2)_{\rm LT}$ is given by the leading twist
expression (\ref{g1partons}) including the target mass
corrections (N=p, n, d).
The dynamical HT corrections $h^N(x)$ in
(\ref{g1F2Rht}) are included
and extracted in a {\it model independent way}. In our analysis
their $Q^2$ dependence is neglected. It is small and the accuracy
of the present data does not allow to determine it. For the
unpolarized structure functions $F_2^N(x,Q^2)_{exp}$ and
$R(x,Q^2)_{exp}$ we have used the NMC parametrization \cite{NMC}
and the SLAC parametrization $\rm R_{1998}$ \cite{R1998},
respectively. The details of our analysis are given in
\cite{LSSHT}. We have found that the fit to the data is
significantly improved when the higher twist corrections to $g_1$
are included in the analysis, especially in the LO QCD case. We
have also found that the size of the HT corrections to $g_1$ is
{\it not} negligible and their shape depends on the target (see
Fig. 2). In Fig. 2 are also presented (open circles) our new
results on the HT corrections to $g_1$ including in the world
data set the recent JLab \cite{JLab} and HERMES \cite{HERMESd}
data. As seen from Fig. 2, the higher twist corrections to the
neutron spin structure functions in the large $x$ region are much
better determined now.
It was also shown (see Fig. 3) that the NLO QCD polarized
PD($g_1^{\rm LT}+\rm
HT$) determined from the data on $g_1$, including higher twist
effects, are in good agreement with the polarized PD($g_1^{\rm
NLO}/F_1^{\rm NLO}$) found earlier from our analysis
\cite{LSS2001} of the data on $g_1/F_1$ and $A_1$ using for the
structure functions $g_1$ and $F_1$ only their {\it leading}
twist expressions in NLO QCD. This observation confirms once more
that the higher twist corrections to $g_1/F_1$ and $A_1$ are
negligible, so that in the analysis of $g_1/F_1$ and $A_1$ data it
is enough to account only for the leading twist of the structure
functions $g_1$ and $F_1$. On the other hand, in fits to the
$g_1$ data themselves the higher twist contribution to $g_1$ must
be taken into account. The latter is especially important for the
LO QCD analysis of the inclusive and SIDIS data.\\

This research was
supported by the JINR-Bulgaria Collaborative Grant, by the RFBR
(No 02-01-00601, 03-02-16816), INTAS 2000 (No 587) and by the
Bulgarian National Science Foundation under Contract Ph-1010.\\

\newpage
\noindent {{\bf Figure Captions}} \vskip 5mm \noindent {\bf
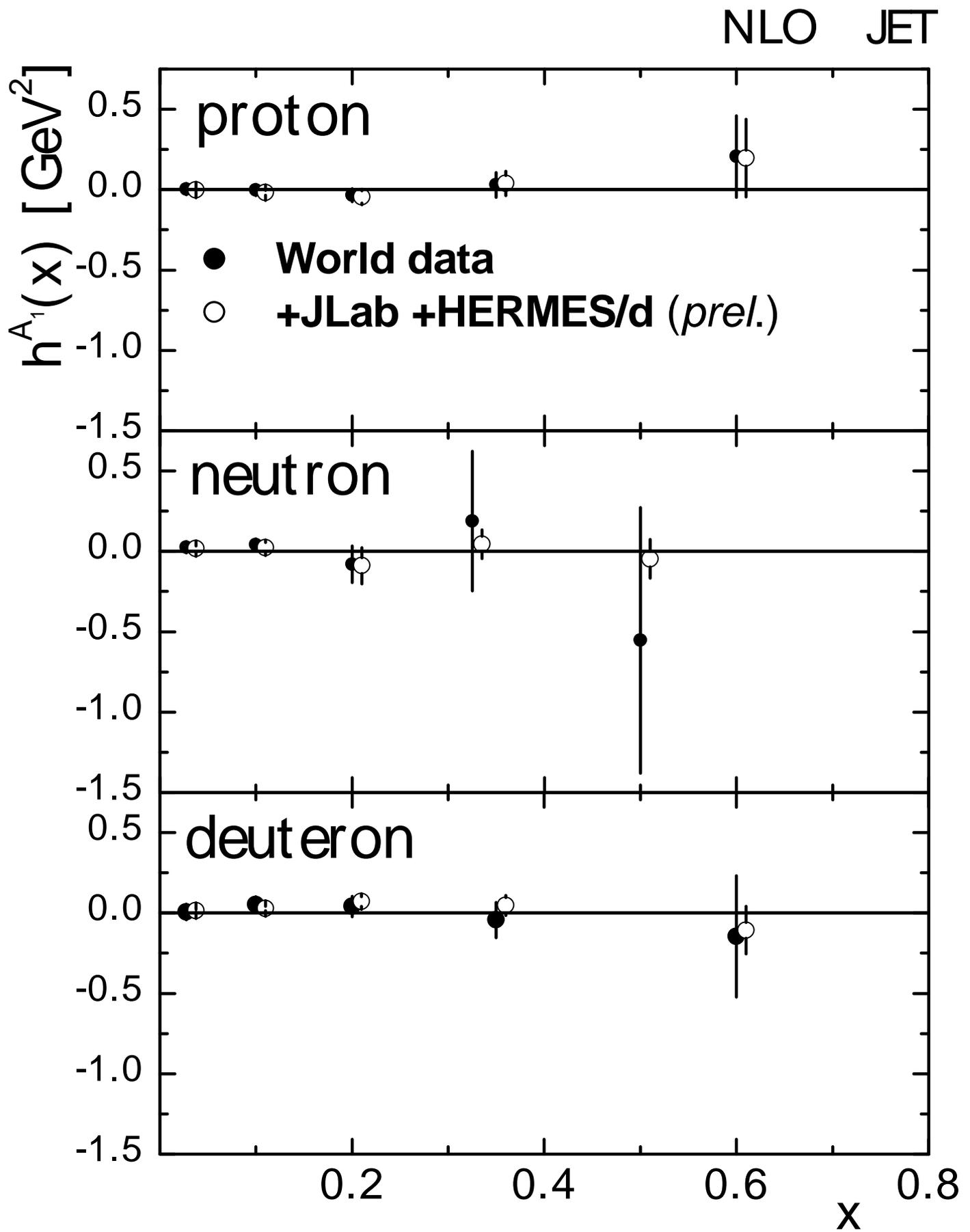
Fig.1.} Effective higher twist contribution $h^{A_1}(x)$ to the
spin asymmetry $A_1^N(x,Q^2)$ extracted from the data.\\

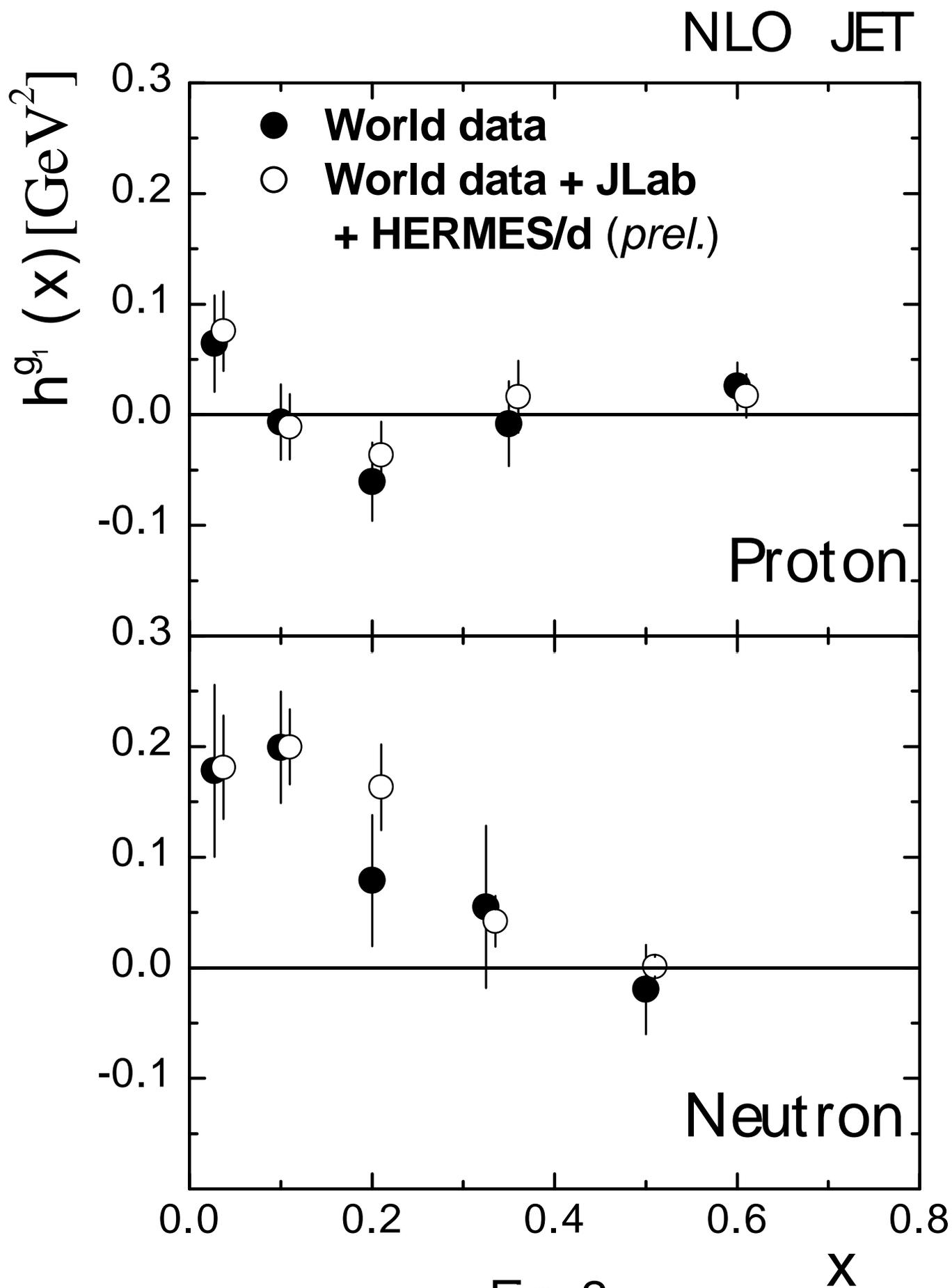
\noindent {\bf Fig. 2.} Higher twist corrections to the proton
and neutron $g_1$ structure functions extracted from the data on
$g_1$ in the NLO QCD approximation for $g_1(x,Q^2)_{\rm LT}$.\\

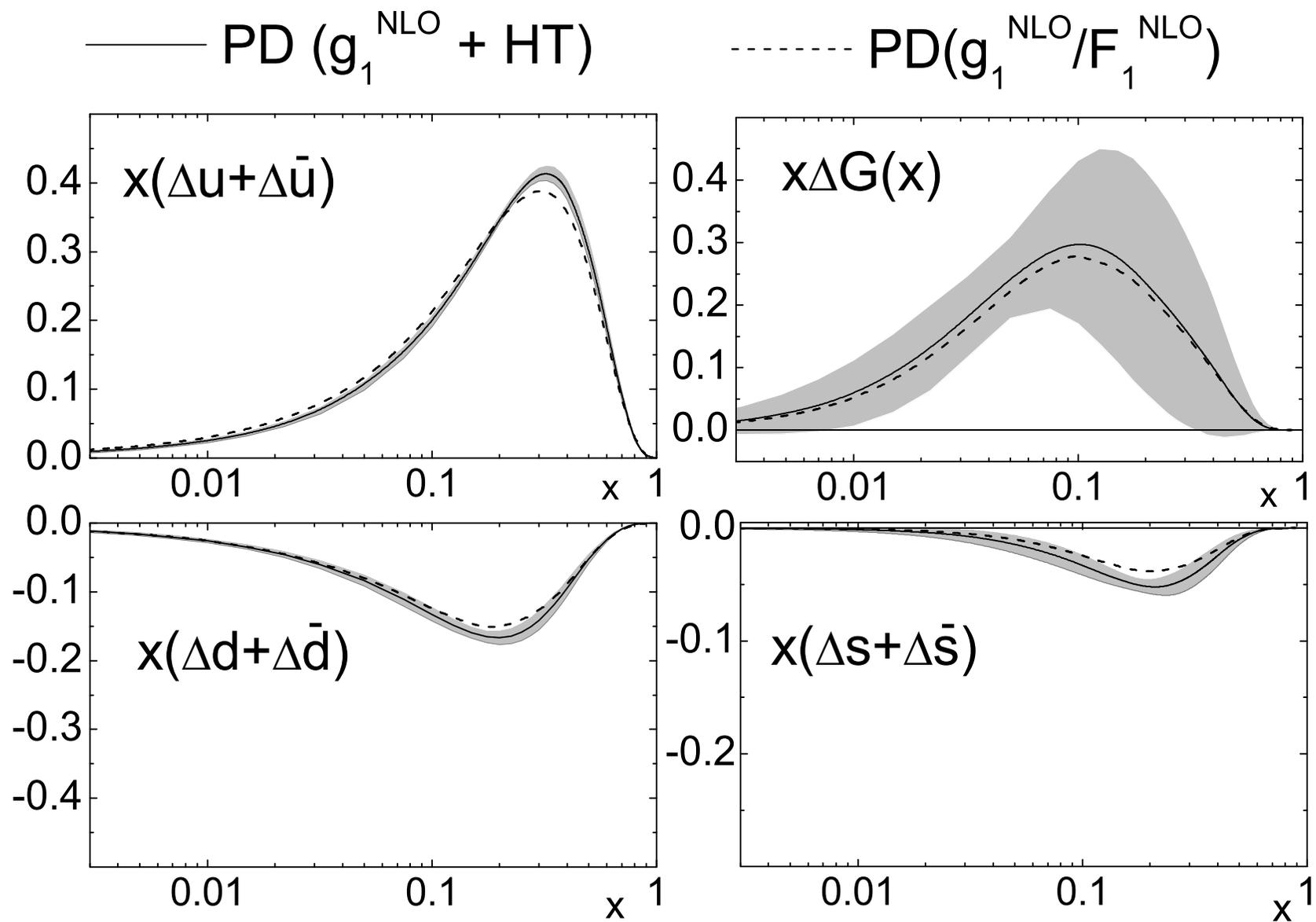
\noindent {\bf Fig. 3.} NLO(JET) polarized parton densities
PD($g_1^{\rm NLO}+\rm HT$) (solid curves) together with error
bands compared to PD($g_1^{\rm NLO}/F_1^{\rm NLO}$) (dashed curves)
at $Q^2=1~GeV^2$. The error bands represent the total errors.

\end{document}